 \documentclass[preprint2]{aastex}

\newcommand{\myemail}{zwchen@pmo.ac.cn}
\usepackage[normalem]{ulem}

\slugcomment{Revised version in Feb 2016}

\shorttitle{Coeval star formation in N4W}
\shortauthors{Z. Chen et al.}

\begin{document}

\title{Coeval intermediate-mass star formation in N4W}

\author{Zhiwei Chen\altaffilmark{1}, Shaobo Zhang\altaffilmark{1}, Miaomiao Zhang\altaffilmark{1}, Zhibo Jiang\altaffilmark{1}}
\affil{$^1$Purple Mountain Observatory $\&$ Key Laboratory for Radio Astronomy, Chinese Academy of Sciences, 2 West Beijing Road, 210008 Nanjing, PR China; \myemail}

\author{Motohide Tamura\altaffilmark{2,3}, Jungmi Kwon\altaffilmark{2,3}}
\affil{$^2$Department of Astronomy, Graduate School of Science, The University of Tokyo, 7-3-1 Hongo, Bunkyo-ku, Tokyo 113-0033, Japan}
\affil{$^3$National Astronomical Observatory of Japan, 2-21-1 Osawa, Mitaka, Tokyo 181-8588, Japan}

\begin{abstract}
Protostars are mostly found in star-forming regions, where the natal molecular gas still remains. In about $5\arcmin$ west of the molecular bubble N4, N4W is identified as a star-forming clump hosting three Class II (IRS\,1\,--\,3), and one Class I (IRS\,4) young stellar objects (YSOs), as well as a submillimeter source SMM1. The near-IR polarization imaging data of N4W reveal two infrared reflection nebulae close to each other, which are in favor of the outflows of IRS\,1 and IRS\,2. The bipolar mid-IR emission centered on IRS\,4 and the arc-like molecular gas shell are lying on the same axis, indicating a bipolar molecular outflow from IRS\,4. There are two dust temperature distributions in N4W. The warmer one is widely distributed and has a temperature $T_\mathrm{d}\gtrsim28\,\mathrm{K}$, with the colder one from the embedded compact submillimeter source SMM1. N4W's mass is estimated to be $\sim2.5\times10^3\,M_\odot$, and the mass of SMM1 is $\sim5.5\times10^2\,M_\odot$ at $T_\mathrm{d}=15\,\mathrm{K}$, calculated from the CO\,$1-0$ emission and $870\,\mu$m dust continuum emission, respectively. Based on the estimates of bolometric luminosity of IRS\,1\,--\,4, these four sources are intermediate-mass YSOs at least. SMM1 is gravitationally bound, and is capable of forming intermediate-mass stars or even possibly massive stars. The co-existence of the IR bright YSOs and the submillimeter source represents potential sequential star formation processes separated by $\sim0.5$\,Myr in N4W. This small age spread implies that the intermediate-mass star formation processes happening in N4W  are almost coeval.

\end{abstract}


\section{Introduction}
Protostars are mostly found in star-forming regions younger than a few Myr. Protostars emit most of their radiations in infrared, and the strong dust extinction along the line-of-sight severely attenuates their optical radiations. Thus protostars are seldom visible in optical wavelengths. Near-IR observations are, in most cases, the important way to study the stellar content of these young star-forming regions. For the circumstellar material of the young stellar objects (YSOs) in these young regions, near-IR and mid-IR observations are always challenging because of the heavy extinction, however, the spatial resolution in near-IR and mid-IR wavelengths are significantly enhanced compared to that in the millimeter and radio wavelength range with the exception of interferometry observation. An effective tool to identify a YSO's outflow in near-IR bands is the imaging polarization observation, which detects the scattered light caused by the dust grain distributed on the cavity blown by the outflow; the so-called polarized infrared reflection nebula (IRN) can be obviously seen in the near-IR polarized images \citep[e.g.,][]{2006ApJ...649L..29T,2007PASJ...59..487K,2011ApJ...741...35K,2012PASJ...64...110C}. In mid-IR bands, dust grain on the cavity wall of an outflow is directly heated by the central YSO, and exhibits strong continuum emission in mid-IR bands \citep[e.g.,][]{2006ApJ...642L..57D,2015A&A...578A..82C}. 

\begin{figure*}[!hbt]
\centering
\epsscale{1.7}
\plotone{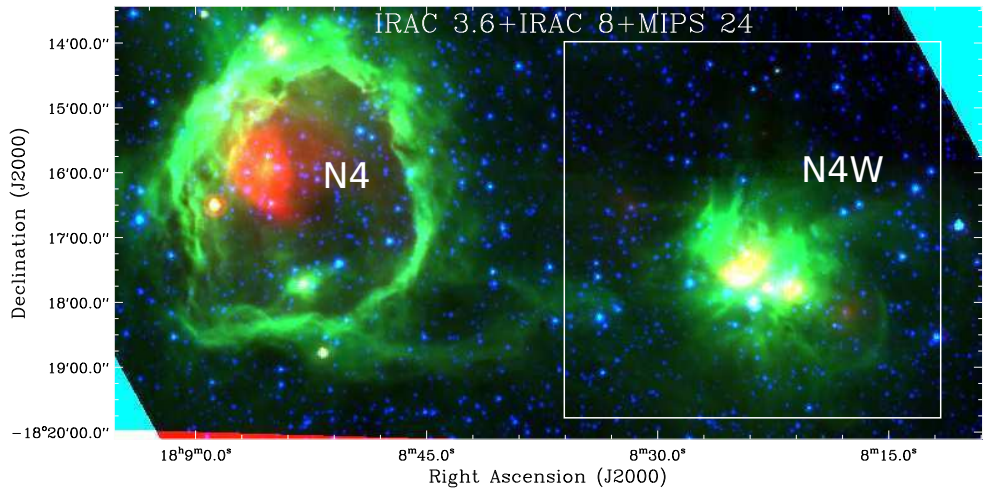}
\caption{Three-color image covering both N4 and N4W, made from the IRAC $3.6\,\mu$m, IRAC $8\,\mu$m, and MIPS $24\,\mu$m data. The white box outlines the field-of-view (FOV) that will be discussed in this paper.}
\label{Fig:Fc}
\end{figure*}

The age spreads of a cluster's stellar population is a good indicator of the overall duration of the star formation process in the cluster. Studies of star-forming regions have reported age spreads range from $<0.1\,\mathrm{Myr}$ to tens of Myr \citep[][and references therein]{2012ApJ...750L..44K}. The age spreads of a cluster could be observationally biased. The young Galactic super star cluster NGC\,3603 might have age spreads of more than 20\,Myr based on the color-magnitude diagrams of the cluster and surrounding area, nevertheless, the age spreads derived from proper-motion selected sample for the core region of NGC\,3603 young cluster are less than 0.1\,Myr \citep[][and references therein]{2014prpl.conf..219S}. This small age spread strongly suggests that star formation in the NGC\,3603 young cluster happened almost instantaneously \citep{2012ApJ...750L..44K}. A reliable age spread estimate for a cluster relies on the proper consideration of the observed region and the secure selection of cluster's members.

The public infrared surveys from near-IR to mid-IR bands are able to detect very young clusters deeply embedded in their natal clouds with spatial resolutions on the order of several arcsecs. These infrared data combined with the submillimeter dust continuum survey and  millimeter CO lines survey enable very detailed investigations for the stellar content, circumstellar material of YSOs, and the overall gas and dust distribution of the molecular clouds. The GLIMPSE survey revealed 322 bubbles that are enhanced in mid-IR bands \citep{2006ApJ...649..759C}. \citet{2010A&A...523A...6D} studied in great detail a smaller sample of 102 bubbles with public data at longer wavelengths. For the bubble N4 contained in both samples, \citet{2013RAA....13..921L} reported CO\,$1-0$ line emission from the molecular ring of N4. The N4's CO\,$1-0$ multi-line data are a part of the Milky Way Imaging
Scroll Painting (MWISP) project which surveys the I\,--\,III quadrants of the Galactic plane with the three CO\,$1-0$ lines \citep[][]{2015ApJ...798L..27S}. In about $5\arcmin$ west of the molecular bubble N4, a molecular clump with similar radial velocity observed in all three CO\,$1-0$ lines is referred to N4W. Unlike N4 that attracts intense interests because of its peculiar morphology, the properties of N4W is left unknown.

This paper presents the results of the near-IR to submillimeter data for N4W. In Section~\ref{sect:Obs} we described the details of near-IR imaging polarization observations, and summarized all the other data used in this paper, which are available from the various public surveys for the Galactic plane. Section~\ref{sect:res} presents the results from these data. We discussed the star formation process in N4W in Section~\ref{sect:dis}, and summarized the conclusions in Section~\ref{sect:conl}.

\section{{Data acquisition}}
\label{sect:Obs}
\subsection{{IRSF/SIRPOL  observations}}
The data were taken with SIRPOL on the 1.4\,m telescope
IRSF at the South African
Astronomical Observatory. SIRPOL is a single-beam polarimeter with an achromatic half-wave plate rotator unit and a polarizer attached to the near-IR camera SIRIUS \citep{2003SPIE.4841..459N,2006SPIE.6269E.159K}. SIRPOL enables wide-field ($\sim 8\arcmin\times8\arcmin$) polarization imaging in the $JHK_s$ bands simultaneously. The observations were made on the nights of 2013 July 8\,--\,9. We made 10\,s exposures at 4 wave-plate angles (in the sequence of $0\degr$, $45\degr$, $22.5\degr$, and $67.5\degr$) at 10 dithered positions. The total integration time is 400\,s per wave-plate angle in all bands. The average full width half maximum of point sources in the $J$, $H$, and $K_s$ bands are $2\farcs0$, $1\farcs9$, and $1\farcs8$, respectively. The data were processed using the pipeline SIRPOL09 developed by Y. Nakajima, including  dark-field subtraction, flat-field correction, median sky subtraction, and frame registration. The products of the pipeline are ready for calculating the Stokes parameters $I$, $U$, $Q$:
\begin{mathletters}
\begin{eqnarray}
 I& = &(I_0+I_{22.5}+I_{45}+I_{67.5})/2 \\
 Q& = & I_0-I_{45} \\
 U& = & I_{22.5}-I_{67.5}
\end{eqnarray}
\end{mathletters}
The counts at the four wave-plate angles are $I_0,I_{22.5},I_{45},I_{67.5}$. Further on, the polarization degree $P$ and polarization angle $\theta$ are derived by 
\begin{mathletters}
\begin{eqnarray}
P&=& \sqrt{U^2+Q^2}/I ~~~~~~~~~~~~~~~\\
\theta&=&\frac{1}{2}\,\arctan(U/Q)~~~~~~~~~~~~~~~
\end{eqnarray}
\end{mathletters}
The polarization angle $\theta$ is counted anticlockwise from the north to the east.

\subsection{Infrared and submillimeter  data from public surveys of the Galaxy}

We retrieved infrared and submillimeter data from the various surveys for the Galaxy. The Galactic Plane Survey (GPS), a part of the UKIDSS Infrared Deep Sky Survey \citep{2007MNRAS.379.1599L}, covers approximately 7000\,deg$^2$ to a depth of $K\sim18$ in the $J$, $H$, and $K$ bands with typical spatial resolution less than $1\arcsec$. The GLIMPSE survey is the Spitzer/IRAC survey for the Galactic plane at $3.6\,\mu$m, $4.5\,\mu$m, $5.8\,\mu$m, and $8.0\,\mu$m with a spatial resolution about $2\farcs5$ \citep{2003PASP..115..953B,2009PASP..121..213C}. The MIPSGAL survey is the Spitzer/MIPS survey for the Galactic plane at $24\,\mu$m and $70\,\mu$m, with spatial resolution about $6\arcsec$ and $18\arcsec$, respectively \citep{2009PASP..121...76C}. The $24\,\mu$m point source catalog based the MIPSGAL survey is released in 2015 Feb \citep{2015AJ....149...64G}. The AKARI/FIS All-Sky Survey Point Source Catalog contains bright far-IR point sources in the Galaxy at $65\,\mu$m, $90\,\mu$m, $140\,\mu$m, and $160\,\mu$m. The flux extraction is made in a circle of $48\arcsec$ radius in all bands \citep{2010yCat.2298....0Y}. ATLASGAL is the APEX Telescope Large Area Survey of the Galaxy at $870\,\mu$m with a beam size of $19\farcs2$ \citep{2009A&A...504..415S}. The ATLASGAL compact source catalog in the Galactic longitude range from $330\degr$ to $21\degr$ is released in 2013 by \citet{2013A&A...549A..45C}.

\begin{deluxetable}{c c c c c c c c c c c c c}
\tabletypesize{\scriptsize}
\centering
\tablecaption{Young stellar objects in N4W \label{Tbl:mag}}
\tablewidth{0pt}
\tablehead{
\colhead{Object} & \colhead{R.A.} & \colhead{Dec.} & \colhead{$J$} & \colhead{$H$} & \colhead{$K$} & \colhead{$3.6\,\mu$m} & \colhead{$4.5\,\mu$m} & \colhead{$5.8\,\mu$m} & \colhead{$8.0\,\mu$m} & \colhead{$24\,\mu$m} & \colhead{$L_\mathrm{IR}$}  & \colhead{Class} \\
           & \colhead{ (J2000)}  & \colhead{(J2000)} & \colhead{[mag]} & \colhead{[mag]} & \colhead{[mag]} & \colhead{[mag]}  & \colhead{[mag]} & \colhead{[mag]} & \colhead{[mag]} & \colhead{[mag]} & \colhead{[$L_\odot$]} &   
}
\startdata
IRS\,1   & 18:08:23.63 & -18:18:03.6 & 12.58  & 11.45 & 9.86 & 7.68 & 6.97 & 6.24 & 5.60 & 4.57\tablenotemark{a} & {\boldmath $1.1\times10^2$}  & II \\
IRS\,2   & 18:08:23.87 & -18:17:59.8 & 16.23 & 12.71 & 10.07 & 7.37 & 6.51 & 6.04 & 5.71 & 3.88\tablenotemark{a} & {\boldmath $1.1\times10^2$}  &  II \\
IRS\,3  & 18:08:23.70 & -18:17:57.2 & 16.48 & 13.81 & 11.45 & 8.68 & 7.69 & 6.76 & 5.76 & 3.69\tablenotemark{a} &  {\boldmath $0.7\times10^2$}  & II \\
IRS\,4  & 18:08:22.86 & -18:17:46.3 & 15.80 & 14.20 &13.02 & $-$ &$-$ &$-$ &$-$ &  2.39\tablenotemark{b} &  $ 1.6\times10^2$ &    I \\
\enddata
\tablecomments{$JHK$ magnitudes are adopted from the DR6 release of UKIDSS/GPS survey. Luminosity is estimated at a distance of 3.14\,kpc. The infrared luminosity $L_\mathrm{IR}$ is obtained by integrating the fluxes between $1-24\,\mu$m.}
\tablenotetext{a}{~IRS\,1\,--\,3 are not fully spatially resolved in the MIPS 24\,$\mu$m image. Hence the $24\,\mu$m fluxes of IRS\,1\,--\,3 were estimated by modeling the MIPS 24\,$\mu$m image with three gaussian point spread functions, and in turn converted to magnitudes in the framework of MIPSGAL point source catalog. The photometric errors of IRS\,1, IRS\,2, and IRS\,3 are estimated to be 0.11, 0.08, 0.07 mag in the 24\,$\mu$m band.}
\tablenotetext{b}{$24\,\mu$m magnitude is adopted from MIPSGAL point source catalog.}
\end{deluxetable}

\begin{figure*}[!hbt]
\centering
\epsscale{1}
\plotone{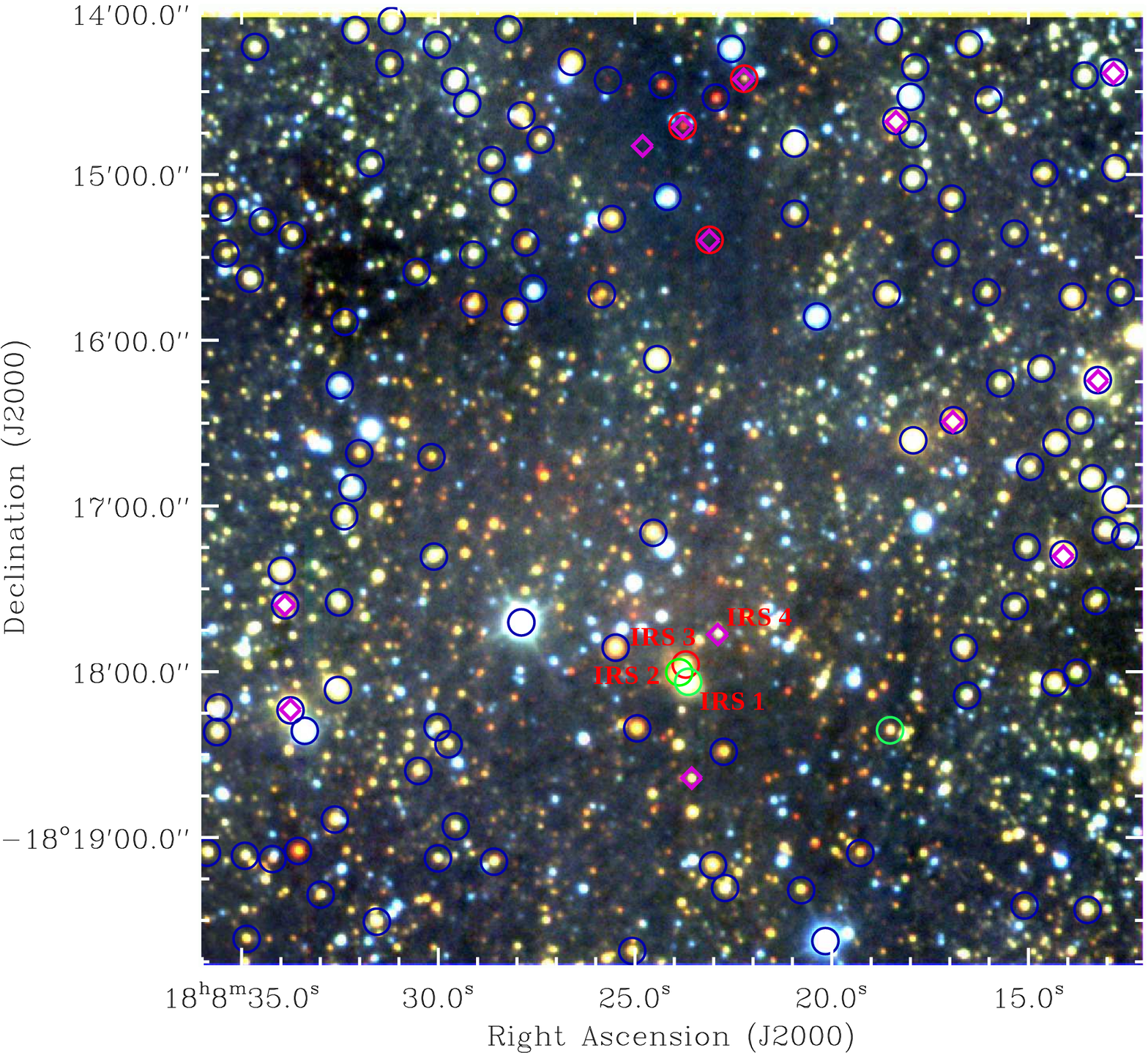}
\plotone{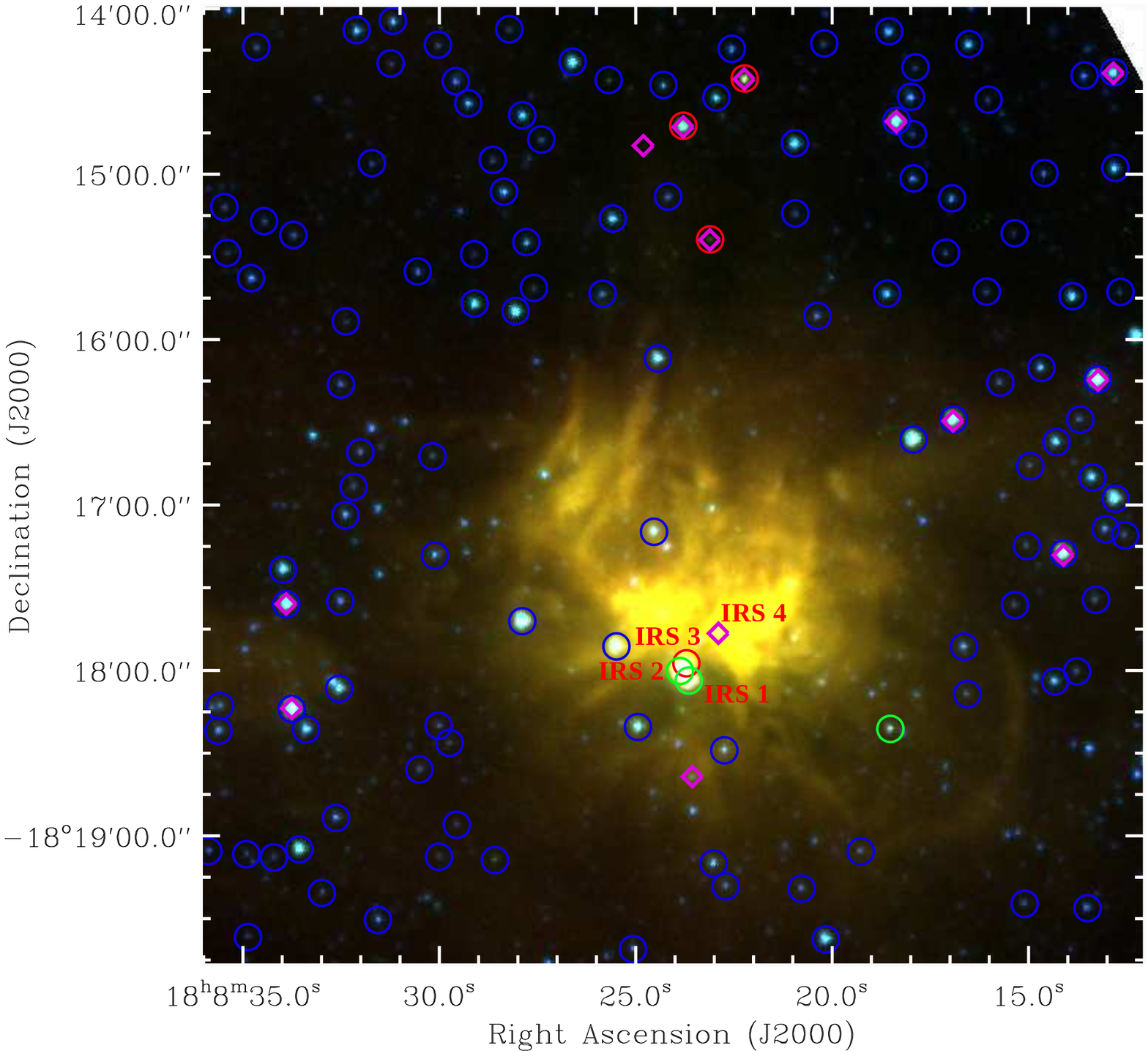}
\caption{Three-color images of N4W, composite of the IRSF $JHK_s$ data (left panel) and IRAC $3.6\,\mu$m, $5.0\,\mu$m and $8\,\mu$m data (right panel). In both panels, blue circles denote Class III, green circles for Class II, red circles for Class 0/I YSOs. The 24$\,\mu$m point sources are labeled by red diamonds. }
\label{Fig:wm}
\end{figure*}

\section{Results}
\label{sect:res}
Fig.~\ref{Fig:Fc} shows the projection of N4W on the sky. It is located only $5\arcmin$ west of the molecular bubble N4. The molecular gas around N4 and N4W shows similar $v_\mathrm{LSR}$ \citep{2013RAA....13..921L}, indicating them physically associated. In addition, we can note an east-west stream prominent in the $8\,\mu$m emission connecting N4 and N4W. Therefore we suggest a distance of 3.14\,kpc for N4W, as the same as N4 \citep{2010A&A...523A...6D}. Both N4 and N4W show prominent $8\,\mu$m emission, which originates from the fluorescent process of PAH molecule excited by far-ultraviolet photons. Different from the ring shape of N4's PAH emission, N4W's PAH emission is compact and concentrated within a radius of $1\arcmin$. In the core region of N4W's PAH emission, two bright point-like sources in $24\,\mu$m are located, as well as two extended objects which are likely warm dust condensations.

Fig.~\ref{Fig:wm} shows the same FOV for the IRSF and Spitzer/IRAC images as the white box in Fig.~\ref{Fig:Fc}. Four bright IR sources are found in the position of N4W, and their photometric data are listed in Table~\ref{Tbl:mag}. IRS\,1--\,3 are spatially resolved in the both IRSF and Spitzer/IRAC images. IRS\,4 is severely contaminated by the mid-IR nebular emission, and misses accurate photometric data in the IRAC bands.

\begin{figure}[!hbt]
\centering
\epsscale{1}
\plotone{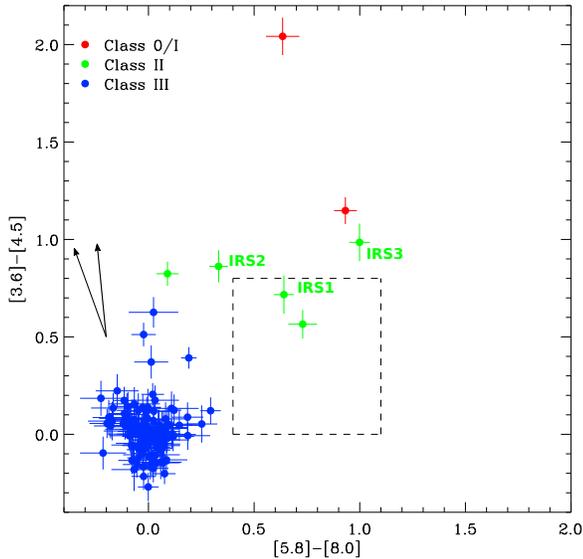}
\caption{IRAC color-color diagram of the 124 point sources with photometric errors $<0.1\,\mathrm{mag}$ in the FOV outlined in Fig.~\ref{Fig:wm}. The rectangle outlined by dashed lines illustrates the domain of Class II, and the domain of Class 0/I is above or right to the Class II domain \citep{2004ApJS..154..367M}. This classification was checked with MIPS $24\,\mu$m photometry; IRS\,2 and IRS\,3 are finally classified as highly reddened Class II in view of their IRAC versus MIPS $24\,\mu$m colors (see more details in Sect.~3.1). We showed reddening vectors for $A_V=30$ derived from the \citet{2001ApJ...548..296W} dust grain models of $R_V=3.1$ (left vector) and $R_V=5.5$ (right vector), respectively.}
\label{Fig:ccd}
\end{figure}

 \subsection{Classification of young stellar objects }
We retrieved the GLIMPSE catalog for the FOV covering N4W. The color-color diagram for the point sources within this region is shown by Fig.~\ref{Fig:ccd}. The color-color diagram based on the four IRAC bands is a promising tool to classify YSOs \citep[e.g.,][]{2004ApJS..154..363A}. Among the 124 point sources with photometric errors less than 0.1\,mag, only seven sources can be classified as YSOs with high significance. Among them the three bright IR sources (IRS\,1\,--\,3) associated with N4W are all YSOs, i.e., IRS\,2\,--\,3 Class 0/I and IRS\,1 a Class II. Three other YSOs are separated from IRS\,1\,--\,3 about $3\arcmin$ in north, and will not be discussed in this paper. 
With the MIPS $24\,\mu$m photometry (see Table~\ref{Tbl:mag}), it is possible to distinguish the highly reddened Class II from Class 0/I YSOs. IRS\,2 is more likely a highly reddened Class II, because its $[5.8]-[24]=2.2$ and $[4.5]-[24]=2.6$ colors both fail with the criteria of Class 0/I ($[5.8]-[24]>4$ and/or $[4.5]-[24]>4$) proposed by \citet{2008ApJ...674..336G}. IRS\,3 has colors of $[5.8]-[24]=3.1$ and $[4.5]-[24]=4.0$; the former one is compatible with that of highly reddened Class II, and the later one is marginally compatible with that of Class 0/I. The colors $[8]-[24]=2.1$ and $[3.6]-[5.8]=1.9$, of IRS\,3, are compatible with the colors of ``hot excess" YSOs, which could be highly reddened intermediate-mass Class II or Class 0/I with more active accretion than typical \citep{2006ApJ...643..965R}. IRS\,3 is more likely a highly reddened Class II than a Class I. Nevertheless, the ambiguity implies that IRS\,3 is during the early time of its Class II stage, just evolved away from Class I.    

IRS\,4 is located in the inner most area of N4W, hence has no entry in the GLIMPSE catalog because of the nebular emission in the $3-8\,\mu$m range. The spectral energy distribution (SED) made from the near- to mid-IR photometric data of IRS\,4 rises much steeper toward longer wavelength than IRS\,1\,--\,2, suggesting a younger PMS stage than Class II, i.e., Class 0/I. Class 0 YSOs are invisible at wavelengths shorter than $2\,\mu$m in most cases because of the optically thick envelope \citep{1993ApJ...406..122A}. In view of IRS\,4's visibility in the $J$ band, we suggest that IRS\,4 is likely a Class I YSO.

The observed IR luminosity, $L_\mathrm{IR}$, integrated between $1-24\,\mu$m is evaluated based on the SEDs of IRS\,1\,--\,4 (see the twelfth column of Table~\ref{Tbl:mag}). Upon the consideration of dust extinction of N4W, the bolometric luminosity $L_\mathrm{bol}$, corrected for extinction effect, must be significantly larger than $L_\mathrm{IR}$. Although IRS\,3's $L_\mathrm{IR}$ is the lowest ($0.7\times10^2\,L_\odot$), IRS\,3 is still much more luminous than most low-mass YSOs, whose $L_\mathrm{bol}$ are mostly far less than $100\,L_\odot$ \citep{2014prpl.conf..195D}. Given the $L_\mathrm{IR}$ of IRS\,1--\,4, we propose that these four sources are at least intermediate-mass YSOs.

\begin{figure}[!hbt]
\centering
\epsscale{1}
\plotone{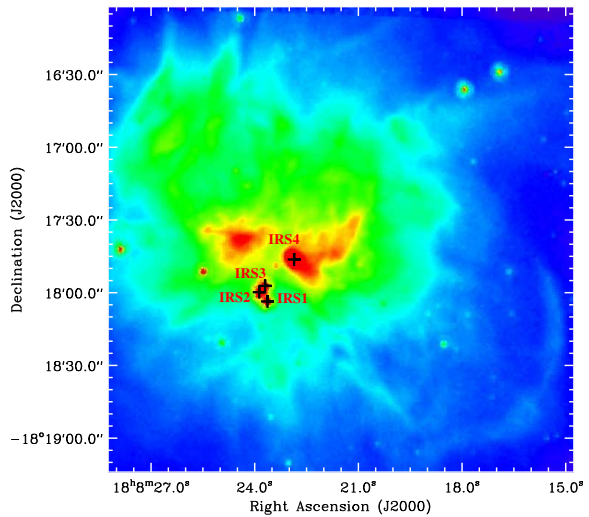}
\caption{Close-up view of N4W in the IRAC $8.0\,\mu$m band. The black crosses mark the positions of IRS\,1\,--\,4.}
\label{Fig:irac4}
\end{figure}

\subsection{Outflow identification}
\begin{figure}[!hbt]
\centering
\epsscale{1}
\plotone{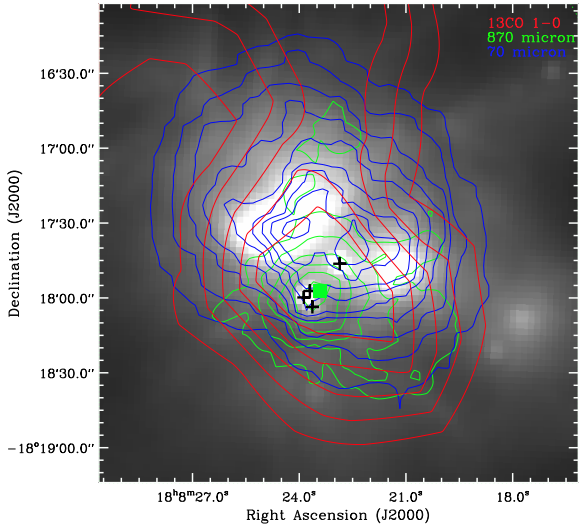}
\caption{Gray-scale intensity map of the MIPSGAL 24$\,\mu$m image for N4W, superposed with the MIPSGAL 70$\,\mu$m contours (blue), ATLASGAL 870\,$\mu$m contours (green), and MWISP $^{13}$CO\,$1-0$ contours (red). Contours start from the half maximum level. The black crosses mark the positions of IRS\,1\,--\,4, and the filled green square marks the position of submillimeter source SMM1. }
\label{Fig:dust}
\end{figure}

The compact mid-IR nebular emission in the close proximity to IRS\,4 seen in the IRAC $5.8\,\mu$m and $8.0\,\mu$m images is interesting (see Fig.~\ref{Fig:irac4} for example). It shows a bipolar shape with two lobes in east and west of IRS\,4. In the lower right corner of Fig.~\ref{Fig:irac4}, we noted an arc-like mid-IR emission located $1\farcm9$ away along the axis of this bipolar structure. The mid-IR bright bipolar, and the mid-IR faint arc-like structure both lie on the axis passing by IRS\,4. The bipolar structure can also be seen in the MIPS $24\,\mu$m image in the form of two warm dust condensations on the two sides of IRS\,4 (see Fig.~\ref{Fig:dust}). In the $24\,\mu$m band, the dust continuum dominates the mid-IR emission. Because the bipolar structure is nearest to IRS\,4, this mid-IR bright structure likely corresponds to a bipolar dust distribution that is directly heated by IRS\,4. The most probable explanation for the alignment between the bipolar and arc-like structure, and IRS\,4 is that both the bipolar and arc-like structure trace the molecular outflow from IRS\,4. The arc-like structure is plausibly a shell of molecular gas that was swept by IRS\,4's outflow to its current position. This molecular gas shell is barely visible in the $24\,\mu$m band. In this case, the fluorescent excitation of the PAH molecules of the molecular gas shell dominates the mid-IR emission in the $5.8\,\mu$m and $8.0\,\mu$m bands. The mid-IR emission of the bipolar structure in the $5.8\,\mu$m and $8.0\,\mu$m bands contains both PAH emission and dust continuum emission. 

\begin{figure}[!hbt]
\centering
\epsscale{1}
\plotone{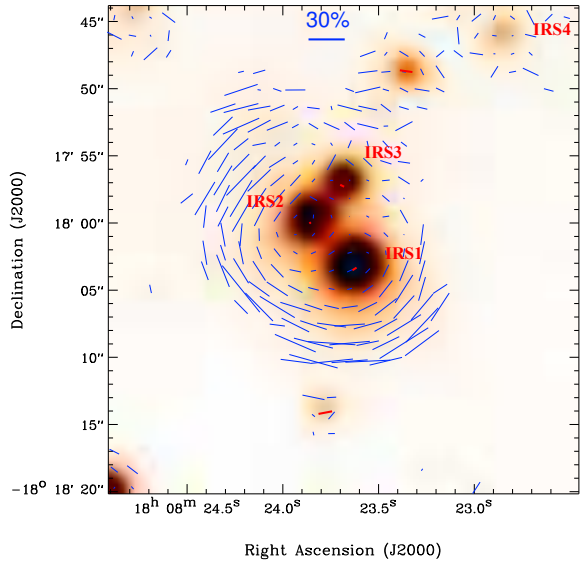}
\caption{IRSF/$K_s$ band polarization vectors of N4W overlaid on the intensity image in the same band. Note that no nebulosity around IRS\,4 is seen in the $K_s$ band, thus only the region around IRS\,1\,--\,3 is closed-up.}
\label{Fig:IRN}
\end{figure}

\begin{figure}[!htp]
\centering
\epsscale{1}
\plotone{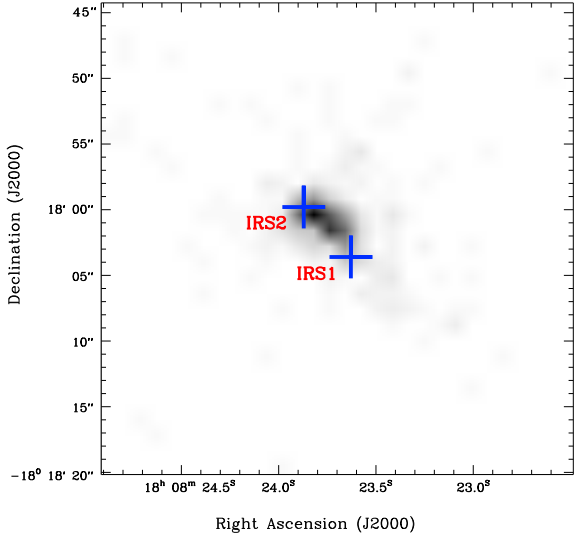}
\caption{Surface density map of the intersection point of any pair of polarization vectors of the IRN associated with IRS\,1\,--\,3. The two blue crosses mark the positions of IRS\,1 and IRS\,2.}
\label{Fig:cen}
\end{figure}

\begin{figure*}[t]
\centering
\epsscale{1.7}
\plotone{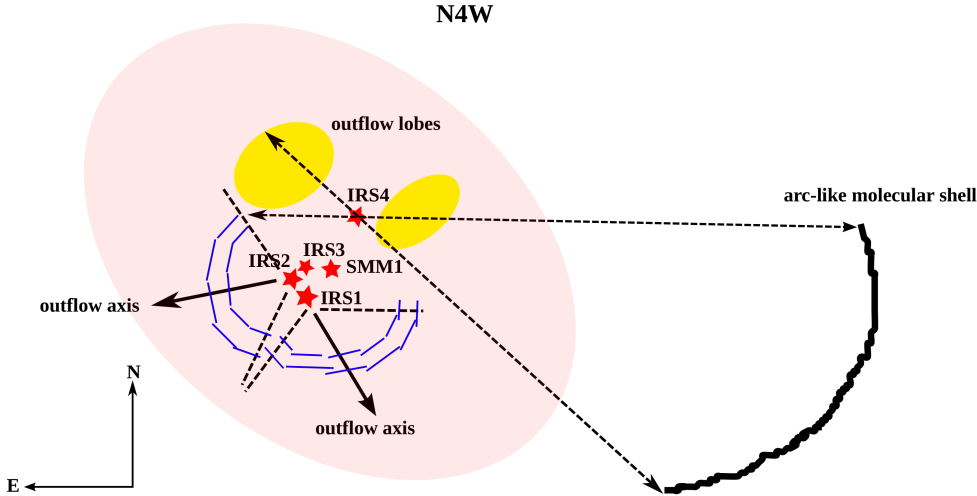}
\caption{Sketch of the star formation activity inside N4W.}
\label{Fig:cartoon}
\end{figure*}

In Fig.~\ref{Fig:IRN} we noted a centrosymmetric polarization pattern in the surroundings of IRS\,1\,--\,3 except the upper-left corner. We regarded this polarized nebula as the IRN associated with IRS\,1\,--\,3. This is the only polarized nebulae found in the surroundings of N4W. The IRN found around IRS\,1\,--\,3 is not surprising because IRS\,1\,--\,3 are the most luminous YSOs in the near-IR bands. The polarization degree of $K_s$-band is around 30\% in the inner area of the IRN, and even exceeds 40\% in the outer area. The larger polarization degree in the outer area also accompanies with larger polarization degree error. The polarization vectors of the IRN associated with IRS\,1\,--\,3 can generally be used to trace back the exact illuminating source. For a singly illuminating IRN, the polarization vectors will mostly intersect at a common position coinciding with the illuminating source if rotating them by $90\degr$. We rotated the polarization vectors of the IRN associated with IRS\,1\,--\,3 by $90\degr$, and showed the intersection point of any pair of polarization vectors in Fig.~\ref{Fig:cen}. Two concentrations of intersection points are clearly found in the close vicinity of IRS\,1 and IRS\,2, indicating two illuminating sources for the IRN. The IRN indeed consists of two smaller IRN; the southern and southwestern part (hereafter IRN\,1) is illuminated by IRS\,1, and the northwestern part (hereafter IRN\,2) is illuminated by IRS\,2. The presence of IRN\,1 and IRN\,2 is in favor of the outflows driven by IRS\,1 and IRS\,2. We could estimate the position angles of the outflows if simply assuming that the cavities blown out by outflows is symmetric with respect to outflow axis. From the polarization patterns of IRN\,1 and IRN\,2, we estimated the open angles to be $125\degr$ and $120\degr$, and the position angles of their symmetric axes to be $30\degr$ and $100\degr$, for IRN\,1 and IRN\,2, respectively (see also Fig.~\ref{Fig:cartoon}).

\subsection{Dust and gas distribution}
\label{Sect:dust+gas}
The H$_2$ column density contours of N4W derived from $^{13}$CO\,$1-0$ emission are compared with the dust continuum emission from  mid-IR to submillimeter (see Fig.~\ref{Fig:dust}). The 24\,$\mu$m emission shows a morphology similar to that of the 70\,$\mu$m emission. Both emission have peak positions very close to the far-IR emission peak of N4W. However, the ATLASGAL 870\,$\mu$m emission shows a point-like morphology with its center coordinating with the peak of H$_2$ column density. The 24/70\,$\mu$m emission is more extended than the 870\,$\mu$m emission. Moreover, the 870\,$\mu$m emission peak \citep[AGAL011.812+00.837 at $18^\mathrm{h}\,08^\mathrm{m}\,23\fs41$, $-18\degr\,17\arcmin\,57\farcs4$;][]{2013A&A...549A..45C} is separated at least $21\arcsec$ from the 70\,$\mu$m peak \citep[at $18^\mathrm{h}\,08^\mathrm{m}\,22\fs9$, $-18\degr\,17\arcmin\,30\arcsec$;][]{2010yCat.2298....0Y}, larger than the beam size of ATLASGAL image. The different properties between 870\,$\mu$m emission and 24/70\,$\mu$m emission indicate two dust condensations of different temperatures; the cold dust condensation is a compact submillimeter source SMM1, and the warm dust spreads over N4W.

\begin{figure}[!hbt]
\centering
\epsscale{1}
\plotone{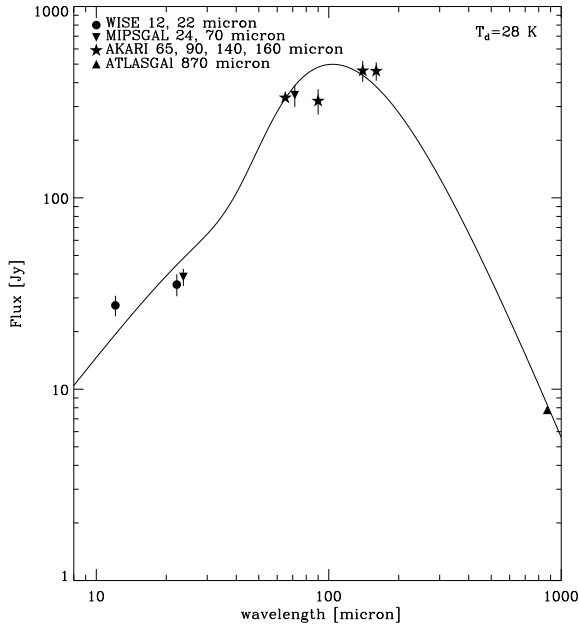}
\caption{SED of the dust continuum emission in N4W. Flux measurements based on the WISE, MIPSGAL and ATLASGAL images are evaluated in a circle of $48\arcsec$ radius centered on the AKARI point source. The SED fitting has made use of the IDL procedure `CMCIRSED' written by \citet{2012MNRAS.425.3094C}.}
\label{Fig:dust_fit}
\end{figure}

The reported fluxes of N4W in the AKARI/FIS catalog comprise both the cold and warm dust continuum emission, because the aperture size of AKARI/FIS catalog is $48\arcsec$. We extracted the fluxes in the same aperture as the AKARI/FIS catalog for the MIPSGAL, WISE, and ATLASGAL images. The dust temperature obtained by fitting the dust continuum emission from 8\,$\mu$m to 870\,$\mu$m is the average value of the warm and cold dust temperatures. The SED fitting suggests a mean dust temperature of 28\,K for the dust in N4W (see Fig.~\ref{Fig:dust_fit}). From the CO observations, the gas temperature of N4W is about 30\,K, very close to the mean dust temperature. Because the resolution of CO observations is comparable to the aperture size of the AKARI/FIS catalog, the gas temperature is most likely the average value of the warm and cold gas. 

We assumed a dust temperature $T_\mathrm{d}\sim15\,\mathrm{K}$ for the submillimeter source SMM1, whose $T_\mathrm{d}$ is constrained to be lower than the mean dust temperature 28\,K. We adopted here the same assumptions as \citet{2014A&A...565A..75C}, i.e., a gas-to-dust ratio $R=100$ and $\kappa_\nu\approx1.85\,\mathrm{cm^2\,g^{-1}}$ at $870\,\mu$m. The flux of SMM1 at $870\,\mu$m can be converted into its mass via the formula
$$
M_\mathrm{gas}=\frac{R\,d^2\,S_\nu}{B_\nu(T_\mathrm{d})\,\kappa_\nu}$$
Substituting the integrated flux of SMM1 \citep[6.51\,Jy;][]{2013A&A...549A..45C} and distance $d=3.14$\,kpc will derive a mass of $5.5\times10^2\,M_\odot$ for SMM1. The molecular gas mass of N4W in a circle of $60\arcsec$ radius independently estimated from the H$_2$ column density contours is $1.4\times10^3\,M_\odot$. The $^{13}$CO\,$1-0$ line is optically thick in the inner most area of N4W, hence a gas mass of $1.4\times10^3\,M_\odot$ is the lower limit of the total gas mass of N4W. In this case, the more rarer C$^{18}$O molecule can better probe the densest region of N4W because the C$^{18}$O line is generally optically thin. The H$_2$ column density calculated from the C$^{18}$O\,$1-0$ line of N4W shows a peak value 1.8 times that from the $^{13}$CO\,$1-0$ line. This difference provides a clue that we probably underestimate the total gas mass of N4W by a factor of 1.8 from the $^{13}$CO\,$1-0$ line. Therefore, we estimate that the gas mass of N4W would be $2.5\times10^3\,M_\odot$.

\begin{table}[t]
\centering
\caption{ $L_\mathrm{bol}$ of YSOs in N4W \label{Tbl:Lbol}}
\begin{tabular}{c c c c }
\tableline
\tableline
Object & $L_\mathrm{bol}$\tablenotemark{1} & $A_V$ & $L_\mathrm{bol}$\tablenotemark{2} \\
  & [$L_\odot$] & [mag] & [$L_\odot$] \\
\hline
IRS\,1 & {\boldmath $1.1\times10^2$} &  5 & {\boldmath $2.0\times10^2$} \\
IRS\,2 & {\boldmath $1.2\times10^2$} &  5 & {\boldmath $1.5\times10^2$} \\ 
IRS\,3 & {\boldmath $0.8\times10^2$} &  5 & {\boldmath $1.0\times10^2$} \\
IRS\,4 & {\boldmath $1.8\times10^2$} &  5 & {\boldmath $2.2\times10^2$} \\
\tableline
\end{tabular}
\tablenotetext{1}{~Bolometric luminosity obtained by integrating the \\
~observed SED in the range $0.02-200\,\mu$m. }
\tablenotetext{2}{~Extinction corrected values of the second column.}
\end{table}

\section{Discussion}
\label{sect:dis}
IRS\,1\,--\,4 are embedded inside the gas reservoir of N4W, which causes strong dust extinction. Extinction of main-sequence stars can be estimated from their near-IR colors because of the well calibrated intrinsic colors of main-sequence stars. However, the near-IR colors of PMS stars are the combination of disk thermal emission and dust reddening. The intrinsic colors of PMS stars are also poorly constrained. Thus the near-IR colors of IRS\,1\,--\,4 are useless for estimating their extinction. Given the distance of N4W (3.14\,kpc), the interstellar reddening caused by interstellar dust along the line-of-sight toward N4W is $\sim5$ if assuming an interstellar reddening efficiency $\approx1.8$\,mag/kpc \citep{2003dge..conf.....W}. 
We used this amount of interstellar extinction to deredden the observed SEDs of IRS\,1\,--\,4, and derived their $L_\mathrm{bol}$ by integrating the extinction-corrected SEDs in the range $0.02-200\,\mu$m \citep[see more details in Sect.~4.5 of][]{2015A&A...578A..82C}. The $L_\mathrm{bol}$ of IRS\,1\,--\,4 derived by this method, given in the fourth column of Table~\ref{Tbl:Lbol}, is the crude lower limit of their real bolometric luminosity.

Although significantly larger than $100\,L_\odot$, the lower boundary of massive YSOs' $L_\mathrm{bol}$ is still not well constrained. From a sample of massive YSOs with $A_V$ values constrained by the near-IR spectra, \citet{2013MNRAS.430.1125C} regarded sources with $L_\mathrm{bol}>5\times10^3\,L_\odot$ as massive YSOs. This criterion of massive YSOs suggests that IRS\,1\,--\,4 are all intermediate-mass YSOs at least.

The mass calculation from only one dust continuum emission measurement is sensitive to the assumed dust temperature. In this paper, we assumed a typical dust temperature of 15\,K, and derived $5.5\times10^2\,M_\odot$ for SMM1. A dust temperature of 10\,K can derive a mass of $1.1\times10^3\,M_\odot$, and a dust temperature of 20\,K instead decreases SMM1's mass down to $3.5\times10^2\,M_\odot$. The upper end of SMM1's mass exceeds the least mass $\sim650\,M_\odot$ for ATLASGAL clumps likely to host massive dense cores and high-mass protostars \citep{2014A&A...565A..75C}. The lower end of SMM1's mass instead suggests intermediate-mass star formation. At the location of the submillimeter source SMM1, we extract the C$^{18}$O\,$1-0$ spectrum to evaluate the 1D line width. Because the source size of SMM1 is similar to the beam size of the C$^{18}$O\,$1-0$ data, the line width of the C$^{18}$O\,$1-0$ spectrum must overestimate SMM1's true line width. Hence we get $\delta v<1.49\,\mathrm{km\,s}^{-1}$ for SMM1. We used this $\delta v$ upper limit to estimate SMM1's virial mass $M_\mathrm{vir}$ to be smaller than $430\,M_{\odot}$, given the SMM1's radius $\sim0.5$\,pc. Obviously SMM1 is gravitationally bound. In view of its uncertain mass, SMM1 will form intermediate-mass or even massive stars. Follow-up high resolution sub-/millimeter observations are able to reveal the internal structure of SMM1, and detect the potential massive dense cores.


N4W has a span of star formation stages from massive dense cores to Class I/II YSOs. The recent comprehensive review of the protostellar evolution derives a protostellar (Class 0+I) lifetime of $\sim0.5$\,Myr for low-mass stars \citep{2014prpl.conf..195D}. The protostellar lifetime could be shorter than $\sim0.5$\,Myr for higher mass stars. Because IRS\,1\,--\,3 are intermediate-mass Class II YSOs, especially that IRS\,3 might be at the early time of Class II stage, we suggest a crude age $\gtrsim 0.5\,\mathrm{Myr}$ for IRS\,1\,--\,3 together. Meanwhile IRS\,4 is an intermediate-mass Class I object, we suggest an age $\lesssim 0.5\,\mathrm{Myr}$ for it. We assumed an average age of 0.5\,Myr for them together. Although the internal structure of SMM1 is still unclear, it is likely that protostellar cores already exist inside SMM1. The co-presence of IRS\,1\,--\,4 and SMM1 reflects a possibility that N4W simultaneously harbors intermediate-mass forming stars and protostellar cores. These two epochs of star formation processes are separated by an age $\sim0.5\,\mathrm{Myr}$, which is assumed as the age spread of N4W. This small age spread implies that the intermediate-mass star formation processes in N4W are almost instantaneous.

\section{Summary and conclusion}
\label{sect:conl}
In this paper we analyzed the near-IR polarization imaging data, and infrared to submillimeter imaging data for a star-forming clump N4W, which has a total mass about $2.5\times10^3\,M_\odot$. In N4W's center a submillimeter source SMM1 has a mass of $5.5\times10^2\,M_\odot$ at $T_\mathrm{d}=15\,\mathrm{K}$, estimated from the dust continuum emission at $870\,\mu$m. Two dust temperature distributions are revealed in N4W. The warm dust spreads all over N4W with a temperature $\gtrsim28$\,K, while the cold dust majorly from SMM1 is embedded. SMM1 is massive enough to maintain itself gravitationally bound, and will be able to form intermediate-mass stars or even potentially massive stars. One Class I (IRS\,4) and three Class II (IRS\,1, IRS\,2, IRS\,3) YSOs are found in the inner most area of N4W. We identified three outflows driven by these YSOs; two of them are traced by the infrared reflection nebulae IRN\,1 and IRN\,2, and the third one is traced by the bipolar mid-IR emission and arc-like molecular gas shell. Fig.~\ref{Fig:cartoon} outlines the three outflows and their corresponding driving sources. For IRS\,1\,--\,4 we derived the crude lower limits of their $L_\mathrm{bol}$, which suggest intermediate masses for them at least. The co-presence of the infrared bright YSOs IRS\,1\,--\,4 and the submillimeter source SMM1 in N4W possibly reflects two epochs of star formation process separated by $\sim0.5$\,Myr. This small age spread implies that the intermediate-mass star formation processes occurring in N4W are almost coeval.

\acknowledgments
This work is supported by the Strategic Priority Research Program `The Emergence of Cosmological Structure' of the Chinese Academy of Sciences, grant No. XDB09000000, the Millimeter Wave Radio Astronomy Database, and the Key Laboratory for Radio Astronomy, CAS. Z.J. acknowledges the support by NSFC 11233007. Z.C. acknowledges the supporting astronomers of the IRSF/SIRPOL observations. This research made use of NASA's Astrophysics Data System Bibliographic Services. This research has made use of the NASA/IPAC Infrared Science Archive, which is operated by the Jet Propulsion Laboratory, California Institute of Technology, under contract with the National Aeronautics and Space Administration.

{\it Facilities:} \facility{IRSF}

\bibliographystyle{apj}
\bibliography{myrefs}
\end{document}